# The origin of the turn-on phenomenon in $T_d$-MoTe$_2$

Q. L. Pei[1†], W. J. Meng[2†], X. Luo[1*], H. Y. Lv[1], F. C. Chen[1,3], W. J. Lu[1],

Y. Y. Han[2], P. Tong[1], W. H. Song[1], Y. B. Hou[2], Q. Y. Lu[2,3,4*], and Y. P. Sun[2,1,4*]

[1] *Key Laboratory of Materials Physics, Institute of Solid State Physics, Chinese Academy of Sciences, Hefei, 230031, China*

[2] *High Magnetic Field Laboratory, Chinese Academy of Sciences, Hefei, 230031, China*

[3] *University of Science and Technology of China, Hefei, 230026, China*

[4] *Collaborative Innovation Center of Advanced Microstructures, Nanjing University, Nanjing, 210093, China*

## Abstract

We did the resistivity and scanning tunneling microscope/spectroscopy (STM/STS) experiments at different temperatures and magnetic fields to investigate the origin of the turn-on (*t-o*) phenomenon of $T_d$-MoTe$_2$. There are two interesting observations. Firstly, magnetoresistance (MR) follows the Kohler's rule scaling: $MR \sim (H/\rho_0)^m$ with $m \approx 1.92$ and the *t-o* temperature $T^*$ under different magnetic fields can also be scaled by $T^* \sim (H-H_c)^\upsilon$ with $\upsilon = 1/2$. Secondly, a combination of compensated electron-hole pockets and a possible electronic structure phase transition induced by the temperature have been validated in $T_d$-MoTe$_2$ by the STM/STS experiments. Compared with the STS of $T_d$-MoTe$_2$ single crystal under $H = 0$, the STS hardly changes even when the applied field is up to 7 T. The origins of the *t-o* phenomenon in $T_d$-MoTe$_2$ are discussed. Meanwhile, we analyzed the universality and applicability of the *t-o* phenomenon in the extreme MR materials with almost balanced hole and electron densities as well as with other systems where the density of hole or electron is in dominant position.

[†]These authors contributed equally to this work.

[*]Corresponding author: xluo@issp.ac.cn, qxl@ustc.edu.cn and ypsun@issp.ac.cn

The magnetoresistance (MR) effect is an important phenomenon that the electrical resistance of a material is changed under a magnetic field. The MR effect is the core of hard drives in computers and of magnetic field sensors in other applications. Due to its promising applications in devices, exploring new materials with large MR has remained at the frontier of material science research.[1-3] The materials with larger MR have attracted much attention in the past twenty years. Besides the giant negative MR (GMR) and colossal negative MR (CMR) found in metallic multilayer films and Mn-based oxide thin films, respectively, extremely large positive MR (XMR) has been observed in graphite, bismuth, and many other compounds such as $PtSn_4$, $Cd_3As_2$, Ta/NbAs, Nb/TaP, LaSb/Bi, $WTe_2$, $MoTe_2$ and so on.[4-14] In particular, much attention has been paid to the recently discovered XMR in $WTe_2$ and $MoTe_2$ because they have two-dimensional (2D) structures and can be more easily exfoliated, which is good for fabricating the hetero-junction of Van der waals type with another 2D material.[15]

There is a unique feature-the turn-on (*t-o*) temperature $T^*$ behavior of the material with XMR effect. When the applied magnetic field is above a certain value $H_c$, the temperature dependent resistivity $\rho(T)$ shows a minimum at $T^*$. As $T < T^*$, the resistivity increases dramatically with the decreasing temperature. When $T > T^*$, it shows a metallic temperature dependence similar to that at zero-field. There are several different interpretations on the origins of the *t-o* phenomenon in XMR materials. For example, in $WTe_2$, a magnetic-field-driven metal-insulator transition is considered as the reason of such a marked up-turn behavior. However, this interpretation has been questioned by several previous studies.[16, 17] Meanwhile, the isostructural material $T_d$-$MoTe_2$ also shows the XMR property and *t-o* phenomenon.[18] In order to explore the origins of the *t-o* phenomenon in the materials with XMR effect, further experiments are needed. In this paper, we did the resistivity (under the magnetic field $H$ = 16 T) and scanning tunneling microscope/spectroscopy (STM/STS) experiments (under $H$ = 7 T) at different temperatures to investigate the origins of the *t-o* phenomenon of $T_d$-$MoTe_2$. Firstly, we found that MR follows the Kohler's rule scaling. Secondly, compared with the STS of $T_d$-$MoTe_2$ single crystal under $H$=0, the STS hardly changes even when the applied field is as high as 7 T. Meanwhile, an electronic structure change induced by the temperature has been suggested in $T_d$-$MoTe_2$ based on the STM experiments, which may be the possible origin of the *t-o* phenomenon in $T_d$-$MoTe_2$. Moreover, we analyzed the universality and applicability of the *t-o* phenomenon of XMR materials with almost balanced hole and electron densities and with other systems where the density of hole or electronics has a dominated position of the electronic properties.

The experimental and calculated details of $T_d$-MoTe$_2$ can be found in the Supporting Information.

Figure 1(a) shows the temperature dependence of the longitudinal resistivity $\rho_{xx}(H)$ of $T_d$-MoTe$_2$ single crystal under different magnetic fields. The resistivity $\rho_{xx}(T,0)$, defined as $\rho_0$, increases with the increase of temperature. The temperature dependent $\rho_{xx}$ at a low magnetic field ($H_c < 2.5$ T) remains a typical metallic behavior from 2 to 300 K. When the magnetic field is raised to 2.5 T or higher, the temperature behavior of the resistivity reaches a minimum at $T^*$, suggesting the possibility of a metal-insulator transition at $T^*$. Based on the prediction by Khveshchenko[19], an external magnetic field can open an excitonic gap in the linear spectrum of the Coulomb interaction quasiparticles in graphite; and with the gap decreases to zero, the temperature follows the relationship $T^* \sim H^{1/2}$, which fits the experimental value of $T^*$ in Fig.1(a). At the same time, we can also analyze the temperature dependent $\rho_{xx}(T,0)$ at low temperatures. According to the Fermi liquid theory: $\rho_{xx}(T,0)=A+BT^2$, where $A = 17.4$ μΩ cm and $B = 0.184$ μΩ cm/K$^2$, the fitting result is expressed as a blue line in the inset of Fig. 1(a). It shows that the experimental data under 70 K is consistent with the Fermi liquid theory.

In order to clarify whether the excitonic gap exists in $T_d$-MoTe$_2$, we scaled the MR curves at different magnetic fields as plotted in Fig. 1(b). We define the MR as follows[20, 21]: $MR=[\rho_{xx}(H)- \rho_{xx}(0)]/\rho_{xx}(0) \times 100\ \%$, where $\rho_{xx}(H)$ is the longitudinal resistivity under applied magnetic field and $\rho_{xx}(0)$ (also presented as $\rho_0$) is longitudinal resistivity under zero magnetic field. The result differs from the expected behavior induced by excitonic gaps in which the MR should change faster at a higher magnetic field.[20] The contradiction between the theory and experiment implies that the origins of the *t-o* temperature behavior in $T_d$-MoTe$_2$ crystals may not be the metal-insulator transition, which is similar to the observed results in WTe$_2$ crystal.[17] In order to present the regular pattern among *MR*, *H*, and *T*, we scaled all the data in Fig. 1(b) onto a straight red line when plotted as $MR \sim H/\rho_0$. It indicates that the temperature dependence of the MR of $T_d$-MoTe$_2$ single crystal follows the Kohler's rule:

$$MR = \alpha(H/\rho_0)^m \qquad (1)$$

where $\alpha=25$ $(\mu\Omega cm/T)^{1.92}$ and $m=1.92$, which are not dependent of the applied magnetic field.

To investigate the law of the *t-o* behavior described by Eq. (1), we plotted the resistivities as well as their difference: $\Delta\rho_{xx} = \rho_{xx}(T, H=8.5\ T) - \rho_{xx}(T, H=0\ T)$ in Fig. 1(c). Eq. (1) can be rewritten as:

$$\rho_{xx}(T,H) = \rho_0 + \alpha H^m / \rho_0^{m-1} \qquad (2)$$

The second term, $\Delta\rho_{xx}$, is inversely proportional to $\rho_0$ when $m = 2$. Because $T^*(H)$ is the temperature corresponding to the resistivity minimum, we can derive $T^*(H)$ by substituting the differential expression into Eq. (1). In Fig. 1(d), we plot the magnetic field dependence of the temperature $T^*$ and $T^{*2}$ of $T_d$-MoTe$_2$. The derived relationship is shown by the red line, which accurately presents the experimental data. The fitting line in Fig. 1(d) also indicates the existence of $H_c$, and the resistivity $\rho_{xx}(T,H)$ has a minimum value at the critical magnetic field. As shown in Fig. 1(d), the $T^*(H)$ can obviously be described as $T^* \sim (H-H_c)^{1/2}$ with the straight blue line. As we discussed in the inset of Fig. 1(a), the resistivity minima occur in the Fermi liquid state. Finally, we can get the value of the critical magnetic field $H_c = 0.5$ T. Furthermore, Fig. 1(a) also predicts a temperature dependence for the minimal resistivity $\rho^*$ at the critical temperature $T^*$. The curve of $\rho_{xx}(T, H)$ follows the law: $\rho^*=[1+(m-1)^{-1}]\rho_0$ with $m = 1.92 \approx 2$. Therefore, we can see that the Kohler's rule described by the Eq. (1) can predict the resistivity minima and the fantastic *t-o* behavior in $T_d$-MoTe$_2$.

To explore the possible reasons of the *t-o* phenomenon in $T_d$-MoTe$_2$ single crystal, we did the STM/STS experiments at low temperatures under the applied magnetic field on the crystals on which were conducted the resistivity measurements. Before doing the STM/STS experiments, we cleaved the crystals again. As shown in Figs. 2 (a) and (b), the red and blue circles correspond to the inner and outer Te atoms. At the same time, the outer and inner Te atoms can be further assigned the brighter and darker features in Fig. 2(c), which shows a representative STM image of the cleaved $T_d$-MoTe$_2$ crystal obtained at $T$=70 K. The STM image in Fig. 2(c) can be further understood by the fast Fourier transform (FFT) analysis. The inset of Fig. 2(c) describes the FFT image of the atomically resolved topography-the blue and red arrows represent the Bragg vectors and the red rectangle corresponds to the first Brillouin zone. Two different types of Bragg vectors can be clearly found from the bright and less bright features in the inset of Fig. 2(c). It seems that there are two different sets of "lattice" existed in $T_d$-MoTe$_2$ single crystal at $T$=70 K. Dramatic changes have been observed when the temperature is down to 7 K, as shown in Fig. 2(d). The FFT image is presented in the inset of Fig. 2(d). It is shown that there is a unique Bragg pattern. In order to compare the differences between STM topographic images obtained at $T$=70 K and 7 K, respectively, the unit cell under 70 K has the lattice constants $a$ = 6.33 Å and $b$ = 3.47 Å. We cannot get the value of $c$ because the STM image can only show us the properties of the surface. As a comparison, the unit cell under 7 K has the lattice constants $a$ = 6.37 Å and $b$ = 3.61 Å. The change of lattice constant $a$ is very small, and the lattice constant $b$ becomes larger when the temperature drops from 70 K to 7 K. The changes of lattice constants agree with the current in Ref. 24, while the lattice constant $a$ does not change obviously and the lattice constant $b$ becomes larger with the reduction of temperature from 300 K to 100 K. As shown in Figs. 2(e) and (f), the light pattern exists at the center of the unique lattice unit cell at 70 K, but shifts to the middle of the edge at 7 K. The shift forms a unique rectangular lattice as we defined above instead of the triangular one at 70 K. There is no structural transition reported in this temperature range for $T_d$-MoTe$_2$.[22] On the other hand, such an electronic structure change may be related to the top Te layer. Due to the distortion of Mo-Te octahedron, from the top view of the Te layer, the inner and outer Te atoms have different energy scales. At low temperature $T$ = 7 K, the distorted Mo-Te octahedron may be energetically unfavorable, so the inner and outer Te atoms cannot be obviously distinguished and the Te atoms become a unique "lattice". At the same time, the observed change cannot be attributed to the charge ordering because there is no gap opening based on the following STM experiments. Recently, Zhou *et al.* reported that there is a sharp change in carrier densities observed between the medium and low temperature ranges, which indicates a possible electronic phase-transition at low temperatures in $T_d$-MoTe$_2$.[23] Meanwhile, Rhodes *et al.* also found that a broad anomaly is observed in the heat capacity around $T$=66 K.[24] It means that the crystallographic structure of $T_d$-MoTe$_2$ evolves upon cooling at low temperatures. Because the STS experiment is very sensitive to the electronic structure of the top layer of the cleaved $T_d$-MoTe$_2$, based on our observations, the distortion of the Te atoms may respond to this electronic phase transition of $T_d$-MoTe$_2$ at low temperatures.

To further study the electronic nature of the $T_d$-MoTe$_2$ single crystal, we performed the STS measurement. Firstly, we calculated the density of states (DOS) as shown in Fig. 3 (a) and of the band structure as shown in the inset of Fig. 3 (b) based on the density functional theory (DFT) calculations (the details can be found in Supporting Information). Compared with the *dI/dV* spectrum in Fig. 3 (b), the theoretical band calculations agree reasonably well to our spectroscopic results. According to Fig. 3 (a), there is no gap around the Fermi energy, which is a necessary condition for the charge density wave (CDW) to form, so CDW may not be the origins of *t-o* behavior of the $T_d$-MoTe$_2$ single crystal. After the theoretical analysis, we took a spectroscopic measurement consisting of 15 differential tunneling conductance spectra (*dI/dV* versus $V_b$) as shown in Fig. 3 (c). The *dI/dV* spectra were acquired within three unit cells (UCs) marked by blue lines in Fig. 3 (d). All curves exhibit similar line shapes without evidence for an energy gap, and each curve in the same UC tends to gather with the others at low Bias below -0.1 V. A dramatic separation of about 0.13 nm between the lateral positions of the maxima for the empty and for the filled states implies a separation of electron/hole (*e/h*) -like channels. At the same time, a normalized particle/hole asymmetry ratio is shown in Fig. 3 (e). The *h/e* ratio varies within a UC along *b* axis, which is perpendicular to Mo chain. As far as we know, a similar microscopic picture of separated *h/e* channels for a double-carrier system has been observed previously for semi-metallic materials in WTe$_2$.[25] Unlike WTe$_2$ with an *e/h* ratio around 1 (perfect *h-e* compensation), the overall *h/e* ratio of $T_d$-MoTe$_2$ within a unit cell always equals to 1.2. Compared with WTe$_2$, the less perfect *h-e* compensation may correspond to the smaller MR in $T_d$-MoTe$_2$. The *h/e* ratio in MoTe$_2$ varies between 1.1 and 1.3. In the half of the unit cell, it is more electron-like with the ratio lower than 1.2; and vice versa in the other half, even though the overall ratio is always greater than 1, MoTe$_2$ is more hole-like. Because the *t-o* phenomenon of $T_d$-MoTe$_2$ occurs when the applied magnetic field is roughly the critical field $H_c$ = 2.5 T, we did the STS measurements of the $T_d$-MoTe$_2$ under the applied magnetic fields up to *H*=7 T at *T*=7 K. As presented in Fig. 3 (f), compared with the STS without the magnetic field, the shapes of STS under the different applied fields do not show much difference. That means the electronic structure of $T_d$-MoTe$_2$ may not show an obvious change under the magnetic field.

We can now try to understand the origins of the *t-o* phenomenon in $T_d$-MoTe$_2$ single crystal. At first, although the Kohler's rule is phenomenological, it was developed to account for the MR in metals where the density of carrier is a constant in the theoretical derivation. Recently, Kohler's rule plots in XMR materials reached agreement with the experimental results, which indicates that the *t-o* phenomenon in XMR materials may originate from the strong temperature dependence of the high mobilities of the charge carriers.[26-28] For the $T_d$-MoTe$_2$ in our study as shown in Fig. 1 (b), the *MR(H)* data also follow Kohler's rule well (Eq. (1)) except that the data are at very low temperatures. Because the parameters $\alpha = 25\ (\mu\Omega cm/T)^{1.92}$ and $m = 1.92$ are independent of magnetic field and temperature, according to the Eq.(2). And the temperature dependence of resistivity at a fixed magnetic field is solely related to $\rho_0$, which is also independent of magnetic field. In the formula $\rho_0=1/ne\mu$, $n$, $e$ and $\mu$ are the carrier density, the electron charge and the carrier mobility, respectively.[29] It means that the $\rho_0$ is inversely proportional to the temperature dependence of the mobilities $\mu_e(T)$ and $\mu_h(T)$ and the carrier densities $n_e(T)$ and $n_h(T)$ in $T_d$-MoTe$_2$. As shown in Fig. 4S, the Hall resistivity measurements reveal that densities and mobilities of *e* and *h* fluctuate as temperature changes. It is different from other XMR materials where the carrier density is bearly varies-only within 10% as temperature changes.[26-28] The *t-o* phenomenon in $T_d$-MoTe$_2$ may originate from the collaborative effect of the strong temperature dependence of the high mobilities and densities of the charge carriers.

Secondly, if a kind of material has perfect *e-h* compensation, the Kohler's rule can be simplified into $MR\sim(H/\rho_0)^2$ according to the two-band model.[30] The value of *m* obtained from Eq.(1) is about 1.92, which is lower than that of a perfect *e-h* compensated system. The results obtained from the Kohler's rule are consistent with the STM observations, as shown in Fig. 3(e), in which the value of *h-e* ratio is about 1.2 at low temperatures, indicating that the perfect *e-h* compensation does not exist in $T_d$-MoTe$_2$. Meanwhile, the detectable disparities between the experimental data at low temperatures and the fitting according to Eq.(1) as shown in Fig. 1(b) are related to the unequality between the densities of the two types of charge carriers in $T_d$-MoTe$_2$. Finally, the STM images of $T_d$-MoTe$_2$ show obvious difference between $T = 70$ and 7 K and the STSs show the undetectable disparities under $H = 0$ and 7 T at $T$=7 K. We can see that the electronic structure of $T_d$-MoTe$_2$ may not be sensitive to the applied magnetic field at low temperatures. According to the Hall resistivity measurements, the densities of hole and electron change significantly at low temperatures and the reformation of the electronic structure occurs around $T \sim 60$ K. MR anisotropy measurements and Hall resistivity data show that an electronic structure change takes place around this temperature[31] As shown in Fig. 4S, as the temperature decreases, the ratio $n_h/n_e$ starts to increase sharply around $T \sim 60$ K and tends to be equal at $T \sim$ 35 K. When *t-o* phenomenon happens, the temperature of the *t-o* phenomenon is about 38 K below $H = 16$ T (as shown in Fig. 1(d)). From Eq.(1), the temperature dependence of resistivity at a fixed magnetic field $\rho_{xx}(T,H)$ is solely related to $\rho_0$, which is independent of the applied magnetic field. For $T_d$-MoTe$_2$, though the carrier densities *n* and mobilities *μ* vary sharply below $T \sim 60$ K, the value of *nμ* changes dramatically only around $T \sim 38$ K where the *t-o* phenomenon presents. In other words, the *e-h* compensation benefits the *t-o* phenomenon in $T_d$-MoTe$_2$ and the change of electronic structure around $T \sim 60$ K is the driving force of the *e-h* compensation. Therefore, it suggests that the *t-o* phenomenon in $T_d$-MoTe$_2$ should be triggered by the electron structure change around $T \sim 60$ K. At the same time, it is noticed that there are some generic features in $T_d$-MoTe$_2$ and WTe$_2$: the XMR effect, similar temperatures of the *t-o* phenomenon, and possible electronic structure change at low temperatures. More importantly, the change of electronic structure happens at low temperature ($T \sim 60$ K for $T_d$-MoTe$_2$ and $T \sim 50$ K for WTe$_2$), which triggers the *t-o* phenomenon of the two compounds. We may expect that the electronic structure change is necessary for the presence of *t-o* phenomenon in WTe$_2$ and $T_d$-MoTe$_2$.

In summary, we have carried out the resistivity and STM/STS measurements at different temperatures and magnetic fields so as to investigate the origins of the *t-o* phenomenon in $T_d$-MoTe$_2$. The experimental data exhibit two interesting results: 1). MR below *t-o* temperature $T^*$ follows Kohler's rule of scaling: $MR \sim (H/\rho_0)^m$ with $m \approx 1.92$; and the $T^*$ for different magnetic fields can also be scaled as $T^* \sim (H-H_c)^\upsilon$ with $\upsilon = 1/2$. 2). Compensated *e-h* pockets and an electronic structure change induced by temperature presented in $T_d$-MoTe$_2$ via STM/STS experiments. Comparing it with the STS under $H = 0$ T, we have found that the STS shapes do not change much even if the applied magnetic field is 7 T. We have also analyzed the universality and applicability of the *t-o* phenomenon in the extreme MR materials with almost balanced hole and electron densities and with other systems where the density of hole or electron is in dominant position.

# Acknowledgements

This work was supported by the National Key Research and Development Program under contracts 2016YFA0300404 and 2016YFA0401003 and the Joint Funds of the National Natural Science Foundation of China and the Chinese Academy of Sciences' Large-Scale Scientific Facility under contracts U1432139 and U1632160, the National Nature Science Foundation of China under contracts 11674326, 11374278, and 51627901, Key Research Program of Frontier Sciences, CAS (QYZDB-SSW-SLH015), the Nature Science Foundation of Anhui Province under contract 1508085ME103 and Hefei Science Center of CAS under contract 2016HSC-IU011. The authors thank Dr. Chen Sun from University of Wisconsin-Madison for her assistance in editing the manuscript.

**Figure 1:**

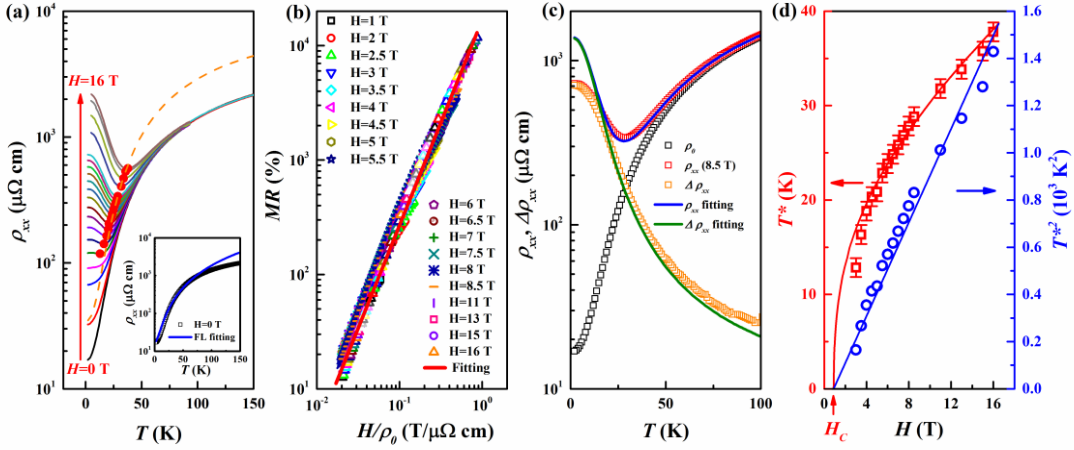

**Fig. 1 (color online): The transport properties of $T_d$-MoTe$_2$.** (a) Temperature dependence of longitudinal resistivity $\rho_{xx}(T)$ of $T_d$-MoTe$_2$ single crystal under different magnetic fields. The upper inset shows the configuration of the measurement. The bottom inset presents the fitting results according to the Fermi liquid theory; (b) The analysis of MR based on the Kohler's rule. The symbols are experimental data and solid line showes a fit to $MR=25(H/\rho_0)^2$; (c) The temperature dependence of the resistivity at H=0 T and 8.5 T and their difference. The solid lines are fitting curves according to Eq.(1) with $\alpha=25[\mu\Omega cm/T]^2$ and $m$=1.92. Since the MR at $T$>100 K is small, we just show the data up to 100 K for clarity; (d) The magnetic field dependent T* for Td-MoTe2 single crystal. The red solid line is the fitting curve according to $\rho_0(T^*)=H[\alpha(m-1)]^{1/m}$ with $m$=1.92 and the dashed blue line represents $T^{*2}$~$(H-0.5)$.

**Figure 2:**

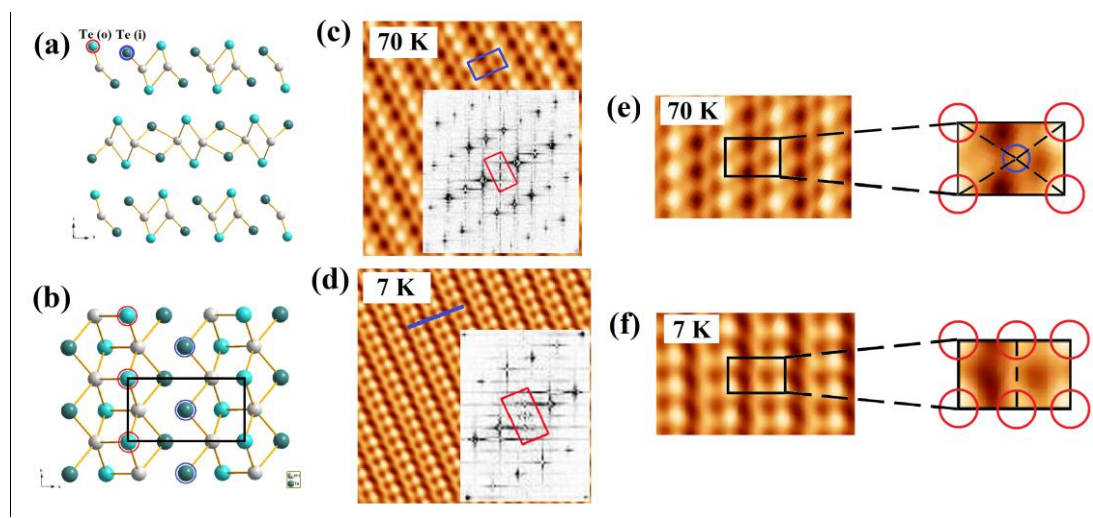

**Fig. 2 (color online): STM/STS measurements of cleaved $T_d$-MoTe$_2$ single crystal surface.** (a) and (b) Side and top view of crystal structure of $T_d$-MoTe$_2$, respectively. Red and blue circles are used for highlighting top Te(o) and Te(i) atoms, which can be observed in atom resolution STM image in (c). The rectangle presents the UC; (c) Atomically resolved STM topographic image of the cleaved $T_d$-MoTe$_2$ surface at T=70 K. The inset shows the FFT image of a high atomic resolution STM topographic image. The red rectangle is corresponding to the first Brillouin zone. Red and blue arrows correspond to the Brag dot with different intensity; (d) Atomically resolved STM topographic image of the cleaved $T_d$-MoTe$_2$ surface at $T$=7 K. The inset shows the FFT image of a high atomic resolution STM topographic image. The red rectangle corresponds to the first Brillouin zone; (e) and (f) The STM difference at $T$=7 K and 70 K at the first Brillouin zone. The light pattern exists at the center of the unique lattice unit cell (shown as a blue circle) at 70 K shifts to the middle of the edge (shown as a red circle) at 7 K.

**Figure 3:**

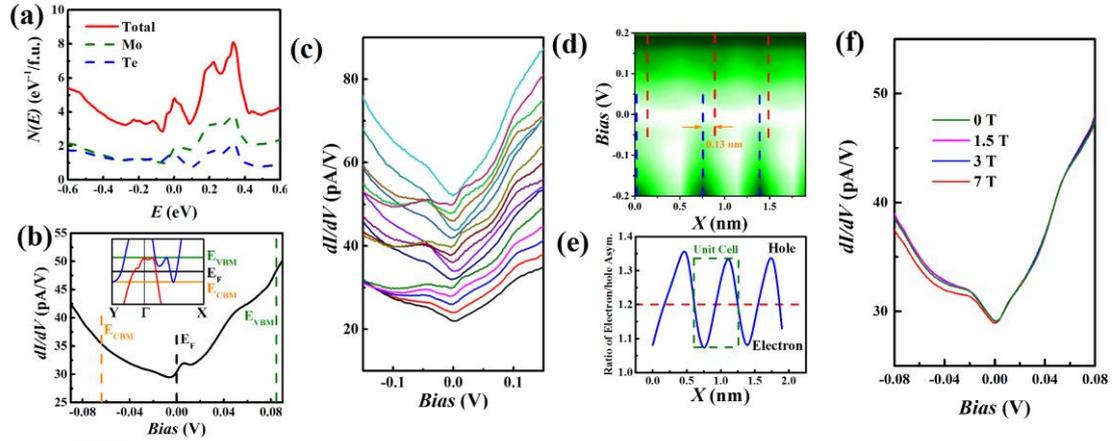

**Fig. 3 (color online): STM image and the differential conductance (*dI/dV*) spectra at *T*=7 K.** **(a)** The calculated total DOS, with projections of the DOS onto Mo (*d* orbitals) and Te (*p* orbitals); **(b)** *dI/dV* spectrum acquired from the surface area in $T_d$-MoTe$_2$ at *T*=7 K. The inset shows the schematic band structure of $T_d$-MoTe$_2$ from the calculations, which can be observed in atom resolution STM image in **(c)**; **(c)** The series of *dI/dV* spectra obtained along the blue line in Fig. 3(d); **(d)** The gray-scale plot of the *dI/dV* curves from (c) with the relative displacement; **(e)** The calculated ratio of *h/e* asymmetry based on the spectra measurement in (a).